\documentclass[10pt,letterpaper,twocolumn, notitlepage]{article}
\usepackage[latin1]{inputenc}
\usepackage{amsmath}
\usepackage{amsfonts}
\usepackage{amssymb}
\usepackage{graphicx}

\usepackage{titlesec}
\titleformat{\section}{\large\bfseries}{\thesection}{1em}{}
\makeatother

\begin{document}

\title{Plasmon Resonances in Nanoparticles, Their Applications to Magnetics and Relation to the Riemann Hypothesis}

\author{I. D. Mayergoyz \\ Department of Electrical and Computer Engineering, UMIACS and AppEl Center, \\ University of Maryland College Park, MD 20742, USA}

\maketitle
\begin{abstract}
The review of the mathematical treatment of plasmon resonances as an eigenvalue problem for specific boundary integral equations is presented and general properties of plasmon spectrum are outlined.  Promising applications of plasmon resonances to magnetics are described.  Interesting relation of eigenvalue treatment of plasmon resonances to the Riemann hypothesis is discussed.
\end{abstract}

\section{Introduction}
\label{}
Plasmon resonances may occur in metallic (gold or silver) nanoparticles under the following two conditions: 1) dielectric permittivity of nanoparticles is negative and 2) the free-space wavelength of incident radiation is appreciably larger than geometric dimensions of nanoparticles.  These two conditions are naturally and simultaneously realized at the nanoscale.  For this reason, plasmon resonances are intrinsically nanoscale phenomena.  These resonances result in powerful nanoscale sources of light which have many important applications in such areas as nano-lithography, near-field optical microscopy, surface enhanced Raman scattering (SERS), biosensors, nanophotonics, optical and magnetic data storage, etc.

In this paper, the eigenvalue approach \cite{ref1}-\cite{ref3} to the analysis of plasmon resonance modes in nanoparticles is reviewed and general properties of plasmon spectrum are presented.  The excitation conditions of desired plasmon resonance modes are discussed along with the time-dynamics of excitation and dephasing of these modes.  Some selective applications of plasmon resonances to magnetics are described.  These applications include plasmon resonance enhancement of magneto-optic effects, all-optical magnetic recording and thermally assisted magnetic recording.  Finally, interesting relation of eigenvalue treatment of plasmon resonances to the Riemann hypothesis is discussed.

\section{Plasmon Resonances as an Eigenvalue Problem}
To start the discussion, we first stress that plasmon resonances are electrostatic in nature, because they occur at free-space wavelengths that are much larger than particle dimensions.  Since the permittivity of nanoparticle is negative, the uniqueness theorem of electrostatics is not valid, and nonzero source-free solutions of electrostatics may appear.  These nonzero source-free solutions are the manifestation of plasmon resonances.  To find these source-free solutions (i.e., plasmon resonance modes), surface electric charges $\sigma(M)$ on the nanoparticle boundary can be introduced (see Figure 1).
\begin{figure}[t]
\centering
\includegraphics{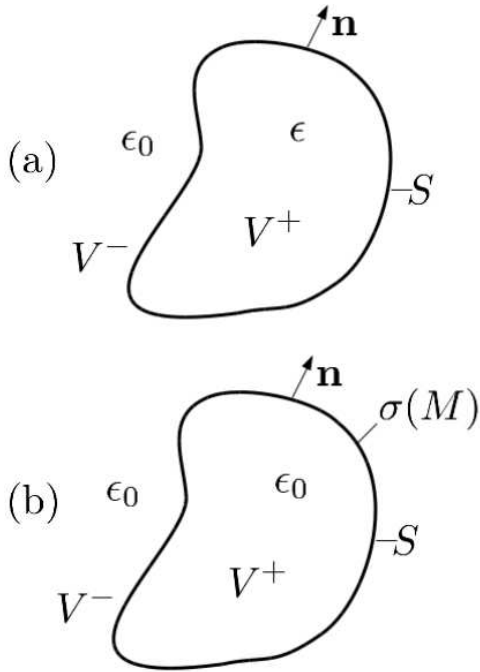}
\caption{(a) Nanoparticle with dielectric permittivity $\epsilon$; (b) Modeling of plasmon resonance modes by using the distribution of surface charges $\sigma(M)$.}
%\vspace{-10pt}
\end{figure}
  The electric field created by these surface charges in free space satisfies all electrostatic equations except the boundary conditions for the normal component of electric field on the nanoparticle surface $S$.  It can be shown [1,2] that the latter boundary condition will be fulfilled if the density of electric charges $\sigma(M)$ satisfies the following boundary integral equation:
\begin{equation}
\sigma(Q) = \frac{\lambda}{2 \pi} \oint_S \sigma(M) \frac{\mathbf{r}_{MQ} \cdot \mathbf{n}_Q}{r_{MQ}^3} d S_M,
\end{equation}
where
\begin{equation}
\lambda = \frac{ \epsilon - \epsilon_0}{\epsilon + \epsilon_0}
\end{equation}
and all other notations have their usual meaning.

It is apparent that nonzero solution of homogeneous integral equation (1) may exist only for special values $\lambda_k$, which are the eigenvalues of the integral operator in (1):
\begin{equation}
\lambda_k = \frac{ \epsilon_k - \epsilon_0 }{ \epsilon_k + \epsilon_0}.
\end{equation}
As soon as eigenvalues $\lambda_k$ are found, the corresponding resonance values of permittivity $\epsilon_k$ can be determined from (3) and then the resonance frequency $\omega_k$ is calculated by using the nanoparticle dispersion relation $\epsilon(\omega)$:
\begin{equation}
\epsilon_k = \epsilon'(\omega_k) = \textrm{Re} \left[ \epsilon(\omega_k) \right].
\end{equation}
The eigenfunctions $\sigma_k(M)$ corresponding to $\lambda_k$ create fields which can be construed as plasmon resonance modes.  

There is another, dual approach to the analysis of plasmon resonance modes.  In this approach, double layers of electric charges $\tau(M)$ are introduced on nanoparticle boundaries instead of single layer of electric charges $\sigma(M)$ (see Figure 2).
\begin{figure}[t]
\centering
\includegraphics{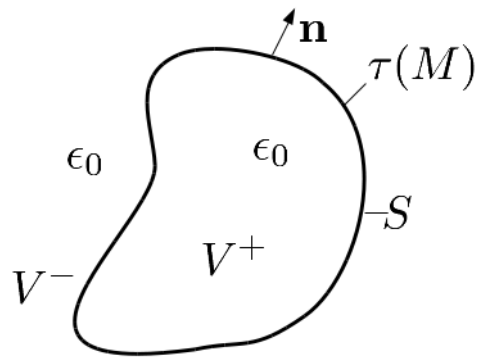}
\caption{Modeling of plasmon resonances by using double layer charges.}
\end{figure}
The distribution of these double layers must be chosen in such a way that they produce in free space the same field of electric displacement $\mathbf{D}$ as may exist for actual plasmon modes in the presence of nanoparticles.  It can be shown (see [1,2]) that the latter will be the case if the density $\tau(M)$ of double layers satisfies the following boundary integral equation:
\begin{equation}
\tau(Q) = \frac{\lambda}{2 \pi} \oint_S \tau(M) \frac{ \mathbf{r}_{QM} \cdot \mathbf{n}_M}{r_{QM}^3} d S_M,
\end{equation}
where $\lambda$ is given by formula (2).  It is apparent that the last integral equation is adjoint to equation (1) and, as discussed below, they have the same set of eigenvalues $\lambda_k$.  The eigenfunctions of equations (1) and (5) are biorthogonal,
\begin{equation}
\oint_S \sigma_k(M) \tau_i (M) d S_M = \delta_{ki}.
\end{equation}
In numerical calculations, boundary integral equations are discretized and reduced to eigenvalue problems for fully populated matrices.  Since these matrices are generated through the discretization of integral operators with $1/r$-type kernels, numerical computations can be considerably sped up by using the fast multipole method [4,5].

It can be shown [6,7] that the spectrum of integral equations (1) and (5) has the following interesting properties.  For any shape of nanoparticles the eigenvalues of integral equations (1) and (5) are real, which implies (according to the Fredholm theory) that these eigenvalues are the same for both equations.  Moreover, $\lambda_0 = 1$ is an eigenvalue, while for all other eigenvalues,
\begin{equation}
|\lambda_k| > 1.
\end{equation}
The eigenvalue $\lambda_0 = 1$ does not belong to the plasmonic spectrum.  Its respective eigenfunction $\sigma_0(M)$ corresponds to the classical Robin problem of distribution of electric charges over the surface of an ideal conductor.  All other eigenvalues correspond to plasmon resonance modes and according to (3) and (7) these modes may exist only for negative $\epsilon_k$, as expected.  It can also be shown that  the eigenvalues $\lambda_k$ are simple poles of resolvents of integral equations (1) and (5), respectively.

Unique spectral properties of plasmon resonances occur for metallic nanowires, i.e., in the two-dimensional case [1-2].  In this two-dimensional case, for any shape of nanowire cross-section the set of eigenvalues $\lambda_k$ ($k \geq 1$) of the corresponding integral equation
\begin{equation}
\sigma(Q) = \frac{\lambda}{\pi} \oint_L \sigma(M) \frac{ \mathbf{r}_{MQ} \cdot \mathbf{n}_Q}{\mathbf{r}_{MQ}^2} dl_M
\end{equation}
consists of pairs of positive and negative eigenvalues of the same absolute value.  This phenomenon of ``symmetric (twin) spectrum'' is due to unique symmetry properties of the mathematical formulation of the plasmon resonance problem which appear in the two-dimensional case due to the existence of the stream function.  The twin-spectrum phenomenon results in two distinct bands of plasmon resonances with relative dielectric permittivities being reciprocals to one another.  This twin-spectrum phenomenon will be used below to establish some relation of plasmon resonances to the Riemann hypothesis.

It is apparent that the mathematical structure of integral equations (1) and (5) is invariant with respect to the scaling of $S$, i.e., scaling of nanoparticle dimensions.  This implies the unique property of plasmon resonances: resonance frequencies depend on particle shape but not particle dimensions, provided that they remain appreciably smaller than the free-space wavelength.

It is clear that the integral operator in equation (1) is not self-adjoint in $L_2(S)$ because the kernel of this operator is not symmetric.  For this reason, the eigenfunctions $\sigma_i(M)$ and $\sigma_k(M)$ corresponding to different eigenvalues $\lambda_i$ and $\lambda_k$ are not orthogonal on $S$ in the usual sense (in $L_2$-norm).  Nevertheless, it can be shown that electric fields $\mathbf{E}_i$ and $\mathbf{E}_k$ corresponding to eigenfunctions $\sigma_i(M)$ and $\sigma_k(M)$ satisfy the following strong orthogonality conditions:
\begin{equation}
\int_{V^+} \mathbf{E}_i \cdot \mathbf{E}_k dv = 0,
\end{equation}
\begin{equation}
\int_{V^-} \mathbf{E}_i \cdot \mathbf{E}_k dv = 0.
\end{equation}
The peculiar feature of these strong orthogonality conditions is that they hold separately in regions $V^+$ and $V^-$.

As previously discussed, plasmon resonances occur when the real part of dielectric permittivity is negative.  This makes the classical formula
\begin{equation}
\tilde{w}_e = \frac{1}{4} \epsilon'(\omega) | \hat{\mathbf{E}} |^2
\end{equation}
for the time-average of stored electric energy density meaningless.  It turns out that the last formula can be properly modified to be valid for slightly lossy (transparent) dispersive media.  The modified formula is
\begin{equation}
\tilde{w}_e = \frac{1}{4} \frac{d [\omega \epsilon'(\omega)]}{d \omega} | \hat{\mathbf{E}} |^2.
\end{equation}

The presented analysis of plasmon modes is purely electrostatic in nature and completely ignores radiation phenomena.  It is shown in [2] that the radiation corrections can be mathematically treated as perturbations with respect to a small parameter which is the ratio of particle dimension (its diameter) to the free-space wavelength.  It is demonstrated that the first-order radiation corrections to the resonance value of dielectric permittivity (hence, to the resonance frequency) and to plasmon mode electric field are equal to zero for any shape of metallic nanoparticle.  The explicit second-order radiation corrections for the resonance values of dielectric permittivity are derived in [2] and it is shown that in the particular case of spherical nanoparticles these second-order radiation corrections coincide with the radiation corrections obtained from the classical Mie theory.

The excitation of plasmon resonance modes by the incident light can be studied by using the biorthogonal expansion of electric charges $\sigma(M,t)$ induced on $S$ during the excitation process:
\begin{equation}
\sigma(M,t) = \sum_{k=1}^\infty a_k(t) \sigma_k(M),
\end{equation}
\begin{equation}
a_k(t) = \oint_S \sigma(M,t) \tau_k(M) dS_M.
\end{equation}
It is clear that the time evolution of the expansion coefficient $a_k(t)$ reveals the time-dynamics of excitation (and dephasing) of the plasmon mode corresponding to the eigenfunction $\sigma_k(M)$.  This dynamics has been studied in [3], and below, only a few, most salient, results from [3] are presented.  Namely, the following analytical expression for the steady state $a_k^{(ss)}(t)$ of the expansion coefficient in (14) can be derived in the case of resonance excitation:
\begin{align}
& a_k^{(ss)}(t)  = \nonumber \\
& -(\mathbf{E}_0  \cdot \mathbf{p}_k) \left[ \frac{ \epsilon' (\omega_k) - \epsilon_0}{\epsilon''(\omega_k)} \cos \omega_k t + \sin \omega_k t \right]. & & 
\end{align}
Here, $\mathbf{E}_0$ is the electric field of the incident radiation, while $\mathbf{p}_k$ is the dipole moment of the $k$-th plasmon mode.  The last expression clearly reveals that the desired plasmon modes are most efficiently excited when the directions of their dipole moments are parallel to $\mathbf{E}_0$ and when $| \epsilon'(\omega_k)| \gg |\epsilon''(\omega_k)|$, where $\epsilon''(\omega_k) = \textrm{Im}[\epsilon(\omega_k)]$.  It is worthwhile to note that the ratio $\epsilon'(\omega_k)/ \epsilon''(\omega_k)$ is most appreciable for gold and silver when the free-space wavelength is between 700 nm and 1100 nm, and this ratio is appreciably higher for silver than for gold.  Thus, as far as the quality of plasmon resonances is concerned, silver is ``gold'' and gold is ``silver''.  This fact has long been appreciated in the SERS research area where silver nanoparticles have been predominantly used in experiments.  For off-resonance excitation with frequency $\omega_0 \neq \omega_k$, the following expression has been derived [3]:
\begin{equation}
a_k^{(ss)} (t) = ( \mathbf{E}_0 \cdot \mathbf{p}_k ) C(\omega_0) \cos (\omega_0 t + \phi),
\end{equation}
where 
\begin{equation}
C(\omega_0) = \sqrt{ \frac{[\epsilon'(\omega_0)-\epsilon_0]^2 + [\epsilon''(\omega_0)]^2}{[\epsilon_k - \epsilon'(\omega_0)]^2 + [\epsilon''(\omega_0)]^2} }.
\end{equation}
The last formula is instrumental for evaluation of the sharpness (width) of plasmon resonances.

In the case of the Drude model for the dispersion of dielectric permittivity, simple analytical expressions for time-dynamics of excitation and dephasing of particular plasmon modes have been derived in [3].

\section{Applications of Plasmon Resonances to Magnetics}
In this section, only a few sample applications of plasmon resonances to magnetics are briefly discussed.  We first consider the enhancement of magneto-optic effects in garnet films by plasmon resonances excited in metallic nanoparticles embedded in these films.  It is known that on the macroscopic level magnetic garnets act as gyrotropic media that discriminate between right-handed and left-handed polarizations of light.  This results in the Faraday magneto-optic effect.  On the microscopic level, magneto-optic effects are controlled by spin-orbit coupling, which is described by the Hamiltonian which depends on local electric fields.  These fields can be optically induced through plasmon resonances in metallic (gold) nanoparticles embedded in garnets, which may eventually lead to the enhancement of magneto-optic effects.  In this way, plasmon resonances can be practically utilized for the enhancement of the Faraday effect as well as for the probing of the origin of this effect on the very fundamental microscopic level.

Some experimental research [8] has been conducted to test the feasibility of the idea described above.  In particular, garnet films have been grown over (100)-oriented substituted gadolinium gallium garnet (SGGG) substrates partially populated with gold nanoparticles.  These gold nanoparticles have been created by evaporating thin layers (about 5 nm) of gold on selected areas of SGGG garnet substrates and subsequently annealing them in air at about 830 $^\circ \mathrm{C}$.  It has been observed that these gold nanoparticles have different dimensions but mostly the same hemispherical shape.  For this reason, they resonate at practically the same frequency (wavelength) due to scale invariance of plasmon resonances.  By using the liquid phase epitaxy (LPE) technique, thin magnetic garnet films of $(\mathrm{Bi},\mathrm{Pr},\mathrm{Y},\mathrm{Gd})_3 (\mathrm{Fe},\mathrm{Gd})_5 \mathrm{O}_{12}$ composition have been grown over gold nanoparticle-populated substrates.  X-ray diffraction measurements have been performed and they reveal that after LPE growth gold nanoparticles survive in crystalline form.  Then, Faraday rotation and magneto-optic loop measurements have been performed and meaningful (non-negligible) enhancement of Faraday rotation which can be attributed to plasmon resonances in embedded nanoparticles has been observed.  
  
Next, we shall briefly discuss the application of plasmon resonances to all-optical magnetic recording.  It has been recently demonstrated [9] that femtosecond magnetization reversals of about 100 $\mu$m spots of magnetic media can be achieved by using only circularly polarized laser pulses.  The direction of magnetization reversals is controlled by the helicity of circularly polarized light.  It is currently perceived that these magnetization reversals occur due to the combination of the local laser heating of magnetic media close to the Curie temperature and the simultaneous action of circularly polarized light as an effective magnetic field with the direction parallel to the light-wave vector.  Furthermore, right- and left-handed circularly polarized light beams act as effective magnetic fields of opposite directions.

The all-optical magnetization switching described above will be technologically feasible for magnetic recording only if the techniques for delivery of nanoscale-focused circularly polarized light are developed.  The latter can be achieved by using optically excited plasmon resonances in uniaxial metallic nanostructures [10-11].  The central point is that in the case of uniaxial metallic nanoparticles (or nano-holes/apertures) the eigenvalues $\lambda_k$ of integral equations (1) and (5) have geometric multiplicity of two and for each $\lambda_k$ there are two identical (up to ninety-degree rotation in space) plasmon eigenmodes.  These two plasmon modes have the same resonance frequency and can be simultaneously excited by the incident circularly polarized light.  These two simultaneously excited plasmon modes form a circularly polarized plasmon mode that can be instrumental for the dramatic nanoscale enhancement (focusing) of the incident circularly polarized light.  It is worthwhile to point out that each of the two above-mentioned plasmon eigenmodes is not rotationally invariant, while their combination as a circularly polarized plasmon mode is rotationally invariant.  It is also apparent that the circularly polarized plasmon modes are most efficiently coupled to the incident circularly polarized light when the direction of propagation of laser light coincides with the symmetry axis of the nanoparticles.

As discussed before, plasmon resonances can be most efficiently excited in the 700-1100 nm free-space wavelength range.  To obtain plasmon resonances in this range, oblate spheroidal nanoparticles with large aspect ratio or nano-rings can be used.  Another alternative is to use spherical or cylindrical nano-shells.  This is because by controlling nano-shell thickness, plasmon resonances can be effectively shifted into the desired wavelength range [12].  Our numerical simulations [10-11] demonstrated that the intensity of incident circularly polarized light can be increased 750 times in the case of silver nano-rings and 550 times in the case of silver nano-shells.

Finally, we shall briefly discuss the possible application of plasmon resonances to thermally assisted magnetic recording (TAMR).  This type of recording has been the focus of considerable technological interest and ongoing research lately.  The central issue is the development of optical sources of nanometer resolution and high light intensity.  The proper profiling of optical spots is also an important practical issue because it may reduce the detrimental effect of the collateral heating of adjacent recorded bits.  It has been demonstrated [13] that plasmon resonances in metallic nanoparticles and perforated metallic nano-films hold unique promise for the development of such optical sources.  In particular, two promising designs have been identified through numerous computations.  They are rectangular bar-type metallic nanoparticles deposited on dielectric substrates and perforated metallic nano-films with rectangular apertures.  The numerical modeling of these structures reveals that plasmon resonances in rectangular bar-type nanoparticles produce somewhat higher light intensity optical spots than the localized plasmon resonances in perforated nano-films.  In addition, it has been found that plasmon resonances in the rectangular bar-type nanoparticles have some advantages as far as proper profiling of optical spots is concerned.

\section{Relation to the Riemann Hypothesis}
The Riemann hypothesis deals with the zeta function which for $\textrm{Re}(s) > 1$ is defined by
\begin{equation}
\zeta(s) = \sum_{n=1}^\infty \frac{1}{n^s} = \prod_i \left( 1 - \frac{1}{p_i^s} \right)^{-1},
\end{equation}
where $p_i$ are prime numbers and the right-hand side of (18) is the celebrated Euler product taken over all prime numbers.

In [14] (its English translation can be found in [15]) Riemann studied the analytical continuation of the zeta function defined by the infinite series in (18).  Riemann found that $\zeta(s)$ extends to the whole complex plane as a meromorphic function with only one simple pole at $s=1$ with residue 1 and derived the functional equation for the extended $\zeta(s)$.  Riemann then introduced the entire function
\begin{equation}
\xi(\lambda) = \frac{1}{2} s (s-1) \pi^{-s/2} \Gamma(s/2) \zeta(s),
\end{equation}
\begin{equation}
s = \frac{1}{2} + i \lambda, \, \, \, (i = \sqrt{-1}),
\end{equation}
where $\Gamma$ is the Euler gamma function.  The functional equation for $\zeta(s)$ implies that $\xi(\lambda)$ is an even function whose zeros have imaginary parts between $-i/2$ and $i/2$.  Then, Riemann conjectured that all zeros of $\xi(\lambda)$ are real.  This conjecture is the Riemann hypothesis.

The zeta function has zeros at negative integers -2, -4, ..., which correspond to the simple poles of the gamma function $\Gamma(s/2)$.  These zeros are referred to as trivial zeros.  The other nontrivial zeros are $s_k = \frac{1}{2} + i \tilde{\lambda}_k$, where $\xi(\tilde{\lambda}_k) = 0$.  These zeros are in the strip $0 \leq \mathrm{Re}(s) \leq 1$ (see Figure 3).  
\begin{figure}
\centering
\includegraphics{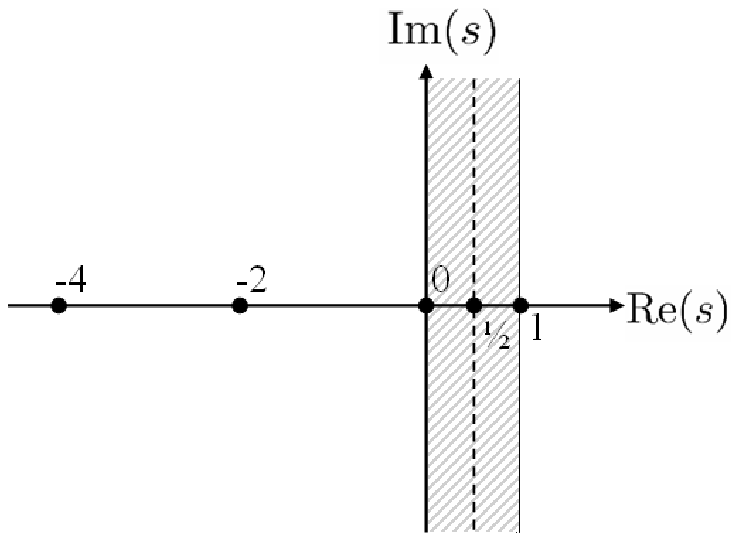}
\caption{The strip $0 \leq \mathrm{Re}(s) \leq 1$.}
\end{figure}
In terms of $s_k$, the Riemann hypothesis is stated as follows: all nontrivial zeros of the zeta function have real part equal to $1/2$.  

Riemann derived the integral representation for $\xi(\lambda)$ and stated that $\xi(\lambda)$ ``can be developed as a power series ... which converges very rapidly.''  This power series can be written as follows [15]:
\begin{equation}
\xi(\lambda) = \sum_{n=0}^\infty \frac{(-1)^n}{2n!} c_{2n} \lambda^{2n},
\end{equation}
where
\begin{equation}
c_{2n} = \int_1^\infty H(x) \left( \ln \sqrt{x} \right)^{2n} dx,
\end{equation}
\begin{equation}
H(x) = 4 \frac{d}{dx}\left[ x^{3/2} \frac{d \psi(x)}{dx} \right] x^{-1/4}
\end{equation}
and
\begin{equation}
\psi(x) = \sum_{n=1}^\infty e^{-n^2 \pi x}.
\end{equation}
It can be proven [15] that for $x \geq 1$
\begin{equation}
H(x) > 0
\end{equation}
and, consequently,
\begin{equation}
c_{2n} > 0.
\end{equation}
It is also apparent from formulas (21)-(24) that $\xi(\lambda)$ assumes real values for real $\lambda$ and the set of its zeros consists of pairs $\pm \tilde{\lambda}_k$, $k \in \mathbb{N}$.  Riemann also suggested the validity of the following product formula,
\begin{equation}
\xi(\lambda) = \xi(0) \prod_k \left( 1 - \frac{\lambda}{\tilde{\lambda}_k} \right) \left( 1 + \frac{\lambda}{\tilde{\lambda}_k} \right),
\end{equation}
which was rigorously proved by Hadamard.

It has been observed in numerical calculations (and then conjectured) that zeros of $\xi(\lambda)$ are simple.

It is our intention to discuss the relation of the Riemann xi-function $\xi(\lambda)$ to the eigenvalue treatment of plasmon resonances in nanoparticles.  Before proceeding with this discussion, it is worthwhile to remark at this point that the Riemann hypothesis stated for xi-function defined by formulas (21)-(24) can be treated as a problem of complex analysis, detached from $\xi$'s number-theoretic origin.  It is in this context that the following two questions can be naturally posed:
\begin{enumerate}
\item  If the Riemann hypothesis is valid, then how is the alignment of all zeros of $\xi(\lambda)$ along the real axis encoded in the mathematical structure of formulas (22)-(24) for the expansion coefficients $c_{2n}$?
\item  What are the necessary and sufficient conditions (in terms of $c_{2n}$) for the alignment along the real axis of all zeros of an entire function defined by the power series (21)?
\end{enumerate}
It is attempted below to answer these questions at least partially.

It has been previously discussed that $\lambda_0 = 1$ is an eigenvalue of the integral equation (8) while all other eigenvalues are real and can be grouped in pairs $\pm \lambda_k$, $k \in \mathbb{N}$ (see Figure 4a).  
\begin{figure}
\centering
\includegraphics{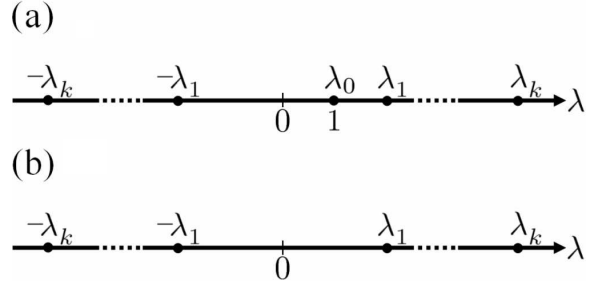}
\caption{(a) Spectrum of integral equation (8); (b) spectrum of integral equation (28).}
\end{figure}
These properties of spectrum of integral equation (8) are valid for any sufficiently smooth curve $L$.  Consider the integral equation 
\begin{equation}
\sigma(Q) = \frac{\lambda}{\pi} \oint_L \sigma(M) \left[ \frac{\mathbf{r}_{MQ} \cdot \mathbf{n}_Q}{r_{MQ}^2} - \frac{\pi}{L} \right] dl_M,
\end{equation}
where $L$ in the integrand stands for the length of curve $L$.  It can be shown that the integral equation (28) has the same spectrum as integral equation (8) except for the eigenvalue $\lambda_0 = 1$ (see Figure 4b).  The proof follows from the facts that for $\lambda_k \neq 1$ we have
\begin{equation}
\oint_L \sigma_k(M) dl_M = 0,
\end{equation}
while for the Robin problem ($\lambda_0 = 1$)
\begin{equation}
\oint_L \sigma_0(M) dl_M \neq 0.
\end{equation}
Indeed, from the well known formula
\begin{equation}
\oint_L \frac{ \mathbf{r}_{MQ} \cdot \mathbf{n}_Q}{r_{MQ}^2} dl_Q = \pi, \quad (M \in L),
\end{equation}
follows the validity of Eq. (29) for the solutions of both integral equations (8) and (28) if $\lambda_k \neq 1$.  Thus, these two equations are equivalent for those $\lambda_k$.  However, this equivalence does not hold for $\lambda_0=1$, because equality (29) is still valid for integral equation (28), while nonzero solution of integral equation (8) exists only under the condition (30).

Now, it is apparent that the Fredholm determinant $D(\lambda)$ of the integral equation (28) is the entire function that assumes real values for real $\lambda$ and the set of its zeros consists of pairs of real numbers $\pm \lambda_k$, $k \in \mathbb{N}$.  As discussed before, these zeros are simple poles of the resolvent of the integral equation (28).  This implies that if $L$ does not have any symmetry and the geometric multiplicity of all $\lambda_k$ is 1 (i.e., no accidental degeneration of $\lambda_k$), then $\lambda_k$ are simple zeros of the Fredholm determinant.  Furthermore, it has been proven (see [16]) that if a kernel of an integral equation is H\"older-continuous with respect to the variable of integration (which is the case for equation (28) when $L$ is smooth) then the Fredholm determinant is of genus zero.  This implies the validity of the product formula for smooth $L$:
\begin{equation}
D(\lambda) = D(0) \prod_k \left( 1 - \frac{\lambda}{\lambda_k} \right) \left( 1 + \frac{\lambda}{\lambda_k} \right).
\end{equation}
Since $D(\lambda)$ is an even entire function, the following power series can be used for its representation:
\begin{equation}
D(\lambda) = \sum_{n=0}^\infty \frac{(-1)^n}{2n!} b_{2n} \lambda^{2n}.
\end{equation}
It is clear from the presented discussion that for various curves $L$ Fredholm determinants of integral equation (28) form the class of entire functions with properties that have been conjectured or proved for the Riemann xi-function $\xi(\lambda)$.  For this reason, the problem can be posed to find such curve $L$ that
\begin{equation}
\xi(\lambda) = D(\lambda),
\end{equation}
which will prove the Riemann hypothesis.  The last formula is consistent with a spectral interpretation of the Riemann hypothesis advanced in the Hilbert-Polya conjecture.  It is worthwhile to remark that this conjecture deals with self-adjoint operator spectral interpretation.  The integral operator in equation (28) is not self-adjoint in the space of square-summable functions because its kernel is not symmetric.  However, this integral operator has real eigenvalues and they are simple poles of its resolvent.  These properties are typical for self-adjoint operators.  For this reason, it is not surprising that this operator will be self-adjoint in properly constructed ``energy'' function space as suggested by the orthogonality conditions (9)-(10).  

Indeed, consider the Hilbert space with the inner product
\begin{equation} 
\left \langle \nu , \sigma \right \rangle = \oint_L \nu (M) \left( \oint_L \sigma (P) \ln  \frac{1}{r_{PM}} dl_P \right ) dl_M
\end{equation}
defined on functions with zero-mean over $L$.  It is easy to see that $\left \langle \sigma, \sigma \right \rangle$ is positive and, up to a factor, it has the physical meaning of energy stored in the electric field created by the charge distribution $\sigma$ over $L$.  By using the relation
\begin{eqnarray}
\oint_L \ln \frac{1}{r_{PQ}} \frac{\partial}{\partial n_Q} \left( \ln \frac{1}{r_{MQ}} \right) dl_Q = \nonumber \\
\oint_L \ln \frac{1}{r_{MQ}} \frac{\partial}{\partial n_Q} \left( \ln \frac{1}{r_{PQ}} \right) dl_Q,
\end{eqnarray}
which follows from the Green's formula, it can be shown that the integral operator in Eq. (28) is self-adjoint in this ``energy'' function space.

From formulas (21), (33)-(34) and the known facts from the theory of integral equations, the following recurrent relations can be deduced:
\begin{equation}
c_{2n} = \oint_L B_{2n-1}(Q,Q) dl_Q,
\end{equation}
\begin{equation}
c_{2n-1}=0,
\end{equation}
\begin{equation}
B_0(Q,M) = K(Q,M),
\end{equation}
\begin{align}
B_n(Q,M) =& \, c_n K(Q,M)   \nonumber \\
 {}& - n\oint_L K(Q,P) B_{n-1}(P,M) dl_P, 
\end{align}
where $K(Q,M)$ is the kernel of integral equation (28),
\begin{equation}
K(Q,M) = \frac{1}{\pi} \left[ \frac{ \mathbf{r}_{MQ} \cdot \mathbf{n}_Q}{r_{MQ}^2} - \frac{\pi}{L} \right].
\end{equation}
It is clear that by using formulas (37)-(41) the iterated traces of the kernel of the integral equation (28)
\begin{equation}
q_{2n} = \oint_L K_{2n}(Q,Q) dl_Q, \, \, \, \, (q_{2n-1}=0),
\end{equation}
\begin{equation}
K_n(Q,M) = \oint_L K(Q,P) K_{n-1} (P,M) dl_P
\end{equation}
can be sequentially found.  

If the kernel (41) can be reconstructed from its iterated traces, then curve $L$ can be reconstructed (up to some scaling) from this kernel.  This is because $\lim_{M \rightarrow Q} \, (\mathbf{r}_{MQ} \cdot \mathbf{n}_Q / r_{MQ}^2)$ is equal to half the curvature of $L$ at $Q$ and a plane curve can be reconstructed from its curvature.

Fredholm determinants can be related to the iterated traces by using another relation from the integral equation theory,
\begin{equation}
-\frac{\xi'(\lambda)}{\lambda \xi(\lambda)} = -\frac{D'(\lambda)}{\lambda D(\lambda)} = \sum_{n=1}^\infty q_{2n} \lambda^{2n-2},
\end{equation}
which is valid for sufficiently small $\lambda$.

The last formula offers another (unrelated to the previous discussion) approach to the pursuit of the proof of the Riemann hypothesis.  Indeed, from the equality 
\begin{equation}
-\frac{\xi'(\lambda)}{\lambda \xi(\lambda)} = \sum_{n=1}^{\infty} q_{2n} \lambda^{2n-2}
\end{equation}
and from the product formula (27), we find
\begin{equation}
\sum_k \frac{1}{\tilde{\lambda}_k} \left( \frac{1}{\tilde{\lambda}_k - \lambda} + \frac{1}{\tilde{\lambda}_k + \lambda} \right) = \sum_{n=1}^\infty q_{2n} \lambda^{2n-2}.
\end{equation}
It is clear that the Riemann hypothesis will be proven if it is established that all poles of the real meromorphic function on the left-hand side of (46) are real.  It is here that the Grommer theorem can be used which gives the necessary and sufficient conditions for the validity of the Riemann hypothesis in terms of expansion coefficients $q_n$.  Namely, according to the Grommer theorem [17,18], the real meromorphic function on the left-hand side of (46) has only real poles, if and only if the expansion coefficients $q_{2n}$ form the positive sequence, i.e., Hankel quadratic forms are positive,
\begin{equation}
\sum_{i,j=0}^N q_{2i+2j+2} \gamma_{2i} \gamma_{2j} > 0
\end{equation}
for any $N$.

To verify inequality (47), a linear functional $f$ defined on the set of polynomials 
\begin{equation}
p(z) = \sum_{i=0}^m p_{2i} z^{2i}
\end{equation}
can be introduced by the formula
\begin{equation}
f(p(z)) = \sum_{i=0}^m q_{2i+2}p_{2i}.
\end{equation}
It is clear that the inequality (47) is satisfied if for any polynomial
\begin{equation}
\gamma(z) = \left( \sum_{i=1}^N \gamma_{2i} z^{2i} \right)^2
\end{equation}
with arbitrary real $\gamma_i$ we have 
\begin{equation}
f\left( \gamma(z) \right) > 0.
\end{equation}
From formulas (21) and (45) the following recurrent relation can be derived: 
\begin{equation}
c_{2n+2} = \sum_{k=0}^n \frac{ (-1)^k (2n+1)!}{ (2n-2k)!} c_{2n-2k} q_{2k+2}.
\end{equation}
By interpreting this relation in terms of the functional $f$ and by using inequalities (26), we find
\begin{equation}
c_{2n+2} = f \left( R_{2n}(z) \right) > 0,
\end{equation}
where
\begin{equation}
R_{2n}(z) = \sum_{k=0}^n r_{2k}^{(2n)} z^{2k},
\end{equation}
\begin{equation}
r_{2k}^{(2n)} = \frac{ (-1)^k (2n+1)!}{ (2n-2k)!} c_{2n-2k}.
\end{equation}
The set of polynomials on which the functional $f$ is positive can be appreciably extended by using the following property of expansion coefficients $c_{2n}$ of $\xi(\lambda)$:
\begin{equation}
\sum_{i,j=0}^N c_{2i+2j+2m} a_{2i} a_{2j} > 0,
\end{equation}
which is valid for arbitrary real $a$'s.  The last property follows from formulas (22) and (25).  By substituting formula (53) into (56) and using linearity of the functional $f$, we find
\begin{equation}
f\left(\sum_{i=0,j=0}^N R_{2i+2j+2m-2}(z) a_{2i} a_{2j} \right) > 0.
\end{equation}
It is not clear at this time if the set of polynomials on which functional $f$ is positive (see inequality (57)) contains all ``complete-square'' polynomials (50).  In other words, it is not clear if the alignment of all zeros of $\xi(\lambda)$ along the real axis is fully encoded in inequalities (25) and (56).  These inequalities do not account (among other things) for the fast decay of $H(x)$ defined by formulas (23)-(24).

The mathematical tools used in the presented discussion of the Riemann hypothesis are quite modest by modern standards.  Nevertheless, it cannot be precluded that these tools may be helpful in the pursuit of the elusive proof of this hypothesis.

This research work has been supported by NSF and ONR.

\end{document}